\documentclass[prb,aps,floatfix,twocolumn,amsmath,amssymb,superscriptaddress]{revtex4-2}
\usepackage{mathtools}
\usepackage{graphicx}\usepackage{bm,braket}
\usepackage[normalem]{ulem}
\usepackage{xcolor}
\usepackage{hyperref}
\hypersetup{colorlinks=true,citecolor=blue,linkcolor=blue,urlcolor=blue}
\DeclareRobustCommand{\erase}{\bgroup\markoverwith{\textcolor{red}{\rule[.5ex]{2pt}{0.4pt}}}\ULon}

\usepackage[T1]{CJKutf8}
\allowdisplaybreaks
\begin{document}
\begin{CJK*}{UTF8}{ipxm}
  \title{Multiple magnetization plateaus induced by farther neighbor interaction in an $S=1$ two-leg Heisenberg spin ladder}

  \author{Hidehiko Kohshiro (幸城秀彦)}
  \affiliation{Institute for Solid State Physics, The University of Tokyo, Kashiwa, Chiba 277-8581, Japan}
  \email{kohshiro@issp.u-tokyo.ac.jp}

  \author{Ryui Kaneko (金子隆威)}
  \affiliation{Department of Physics, Kindai University, Higashi-Osaka, Osaka 577-8502, Japan}
  \affiliation{Institute for Solid State Physics, The University of Tokyo, Kashiwa, Chiba 277-8581, Japan}

  \author{Satoshi Morita (森田悟史)}
  \affiliation{Institute for Solid State Physics, The University of Tokyo, Kashiwa, Chiba 277-8581, Japan}

  \author{Hosho Katsura (桂法称)}
  \affiliation{Department of Physics, The University of Tokyo, Bunkyo-ku, Tokyo 113-0033, Japan}
  \affiliation{Institute for Physics of Intelligence, The University of Tokyo, Bunkyo-ku, Tokyo 113-0033, Japan}
  \affiliation{Trans-scale Quantum Science Institute, The University of Tokyo, Bunkyo-ku, Tokyo 113-0033, Japan}

  \author{Naoki Kawashima (川島直輝)}
  \affiliation{Institute for Solid State Physics, The University of Tokyo, Kashiwa, Chiba 277-8581, Japan}
  \affiliation{Trans-scale Quantum Science Institute, The University of Tokyo, Bunkyo-ku, Tokyo 113-0033, Japan}

  \date{\today}

  \begin{abstract}
    We study the magnetization process of the $S=1$ Heisenberg model on a two-leg ladder with farther neighbor spin-exchange interaction.
    We consider the interaction that couples up to the next-nearest neighbor rungs and find an exactly solvable regime where the ground states become product states.
    The next-nearest neighbor interaction tends to stabilize magnetization plateaus at multiples of 1/6.
    In most of the exactly solvable regime, a single magnetization curve shows two series of plateaus with different periodicities.
  \end{abstract}

  \maketitle
\end{CJK*}
\section{Introduction}
\label{sec:intro}
Quantum fluctuations play a crucial role in low-dimensional systems.
One-dimensional systems have attracted much interest because exact eigenstates are sometimes available, and the ground state can be determined numerically with high accuracy using methods based on the matrix product states~\cite{Takahashi, KBI, Schollwoeck}.
In contrast, the number of exact solutions for two-dimensional quantum spin systems is limited except for some models such as the Shastry-Sutherland model~\cite{SS} and the Kitaev honeycomb model~\cite{Kitaev}.

Quasi-one-dimensional systems, such as zigzag chain and ladder systems, are one-dimensional systems with farther neighbor interactions~\cite{Giamarchi}.
Such interactions can induce geometrical frustration leading to nontrivial phenomena such as magnetization plateaus.
Quasi-one-dimensional systems interpolate one-dimensional and two-dimensional systems and can incorporate geometrical frustration with ease of use.

When an excitation gap opens in the system with $\mathrm{U(1)}$ symmetry under a magnetic field, magnetic plateaus appear.
In general, to induce magnetization plateaus, $p(S-m)$ must be an integer, where $p$, $S$, and $m$ are the period of the ground state, spin, and magnetization per site, respectively~\cite{OYA}.
This formula shows that a large spin $S$ can give rise to different series of magnetization plateaus at multiples of small fractional numbers with periodicity varying.

The effects of farther neighbor interaction have been investigated in spin-ladder systems.
For $S=1/2$, Sugimoto \textit{et al.} found magnetization plateaus at $1/3$, $1/2$, and $2/3$ with the strong rung coupling~\cite{Sugimoto1, Sugimoto2}.
For $S\geq1$, Michaud \textit{et al.} revealed that $i/(4S)$ ($i=0,\ \dots,\ 4S$) plateaus appear in a frustrated spin-$S$ ladder~\cite{Michaud}.
Chandra and Surendran also found a solvable $n$-leg spin-$S$ ladder that produces $i/(2nS)$ ($i=0,\ \dots,\ 2nS$) plateaus~\cite{ChandraSurendran}.

Geometrically frustrated $S=1$ two-leg ladders have been extensively studied motivated by spin-ladder compounds~\cite{OKazaki1, Okamoto1, Okamoto2, Okazaki2, Sakai1, Sakai2, Sakai3, Sakai4, Strecka1}.
A candidate material that exhibits clear magnetization plateaus is an organic antiferromagnet $3$,$3'$,$5$,$5'$-tetrakis($N$-$tert$-butylaminoxyl)biphenyl (BIP-TENO).
It has been considered to be described by the $S=1$ Heisenberg model on a two-leg ladder~\cite{Katoh1, Ohta}.
In the recent experiment of BIP-TENO, $1/4$, $1/3$, and $1/2$ plateaus have been observed~\cite{Nomura}.
The periods of the $1/4$ and $1/3$ plateau states are expected to be two and three, respectively.
It is unusual to observe such magnetization plateaus, whose periods are coprime, simultaneously.

As far as we know, there is no $S=1$ two-leg ladder model where different series of magnetization plateaus, such as $1/4$ and $1/3$ plateaus, coexist.
In most previous theoretical studies for $S=1$ two-leg ladder, the plateaus appear only at multiples of $1/4$~\cite{Michaud, ChandraSurendran, OKazaki1, Okamoto1, Okamoto2, Okazaki2, Sakai1, Sakai2, Sakai3, Sakai4, Strecka1}.
The $1/3$ plateau with the classical up-up-down state has been reported in a spin-$S\geq1$ zigzag ladder~\cite{Dagotto}.
However, there is no $1/4$ plateau in this model.

Considering farther neighbor interactions in the $S=1$ two-leg Heisenberg ladder, we have constructed the model that holds plateaus at multiples of $1/6$ including $1/3$ plateau.
The obtained plateau state is a product state of dimers and different from the classical up-up-down state~\cite{Dagotto}.
We have also found the magnetization curves that hold magnetization plateaus at multiples of $1/4$ and $1/6$ in a certain parameter region where interaction couples up to the next-nearest neighbor rungs.

The paper is organized as follows.
In Sec.~\ref{sec:model}, we introduce the extended Gelfand ladder, the $S=1$ Heisenberg model on a two-leg ladder with farther neighbor two-body spin-exchange interaction that couples nearest and next-nearest neighbor rungs.
We briefly review that the plateaus at multiples of $1/4$ appear when the interaction between next-nearest neighbor rungs is absent.
In Sec.~\ref{sec:onesixth}, we show that the plateaus at multiples of $1/6$ solely appear when the next-nearest neighbor interaction takes a certain value.
In Sec.~\ref{sec:pd}, we calculate energies of the product states and show that our model can realize both $1/6$ and $1/4$ series of plateaus in the same magnetization curve.
In Sec.~\ref{sec:ff}, we prove that the product states used in Sec.~\ref{sec:pd} become the exact ground states of the extended Gelfand ladder in a specific parameter region.
Finally, we summarize our results and give some discussions in Sec.~\ref{sec:sum}.

\section{Model}\label{sec:model}
\begin{figure}
  \includegraphics[width=\columnwidth]{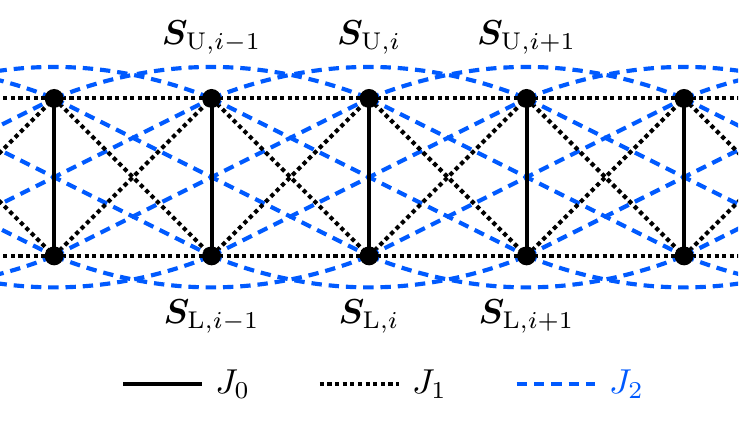}
  \caption{Description of the extended Gelfand ladder.
    We consider Heisenberg-type isotropic exchange interactions.
    Coupling constants along the leg and diagonal directions are the same strength.
  }
  \label{fig:model}
\end{figure}

\begin{figure}
  \includegraphics[width=\columnwidth]{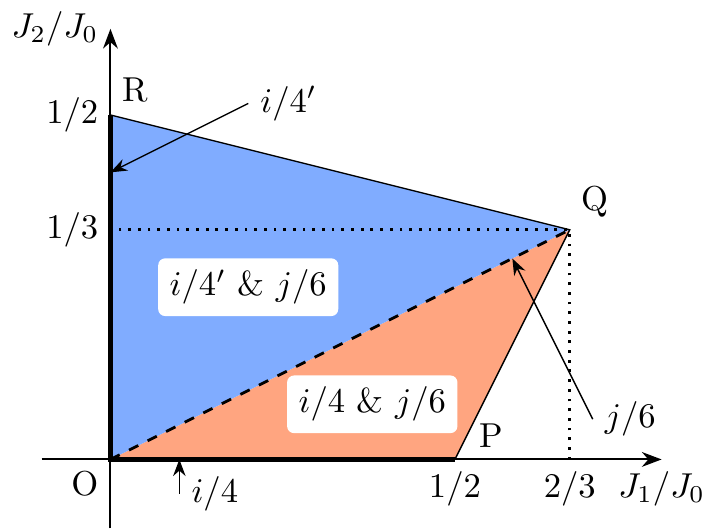}
  \caption{
    $J_1$-$J_2$ phase diagram.
    The colored region is composed of two triangles divided by the dashed line $J_2=J_1/2$.
    Both $1/6$ and $1/4$ series of plateaus coexist in the whole colored region except for three lines $J_1=0, J_2=0$, and $J_2=J_1/2$.
    Only the $1/4$ series ($i/4$) and $1/6$ series ($j/6$) of plateaus appear on the solid line $J_2=0$ and on the dashed line $J_2=J_1/2$, respectively.
    Another $1/4$ series ($i/4'$) of plateaus with period four appear on the solid line $J_1=0$.
    The colored region corresponds to Eq.~\eqref{eq:pos_cond} where the exact ground state is a product state of the highest-weight state.
  }
  \label{fig:region}
\end{figure}

We consider the $S=1$ Heisenberg model under the external magnetic field on a two-leg ladder as depicted in Fig.~\ref{fig:model}.
The Hamiltonian, which we call the extended Gelfand ladder in what follows, under the periodic boundary condition is given as
\begin{multline}
  \mathcal{H}_{\mathrm{full}}
  \coloneqq J_0\sum_{i=1}^L\bm{S}_{\mathrm{U}, i}\cdot\bm{S}_{\mathrm{L}, i}\\
  +J_1\sum_{i=1}^L(\bm{S}_{\mathrm{U}, i}+\bm{S}_{\mathrm{L}, i})
  \cdot(\bm{S}_{\mathrm{U}, i+1}+\bm{S}_{\mathrm{L}, i+1})\\
  +J_2\sum_{i=1}^{L}(\bm{S}_{\mathrm{U}, i}+\bm{S}_{\mathrm{L}, i})
  \cdot(\bm{S}_{\mathrm{U}, i+2}+\bm{S}_{\mathrm{L}, i+2})\\
  -h\sum_{i=1}^L(S^z_{\mathrm{U}, i}+S^z_{\mathrm{L}, i}),
  \label{eq:ham}
\end{multline}
where a subscript U (L) specifies the upper (lower) leg and $i$ is an index along legs.
The site $L+1$ ($L+2$) is identified with $1$ ($2$).

We can rewrite the Hamiltonian by introducing a new $\mathrm{SU}(2)$ spin operator $\bm{T}_i \coloneqq \bm{S}_{\mathrm{U}, i}+\bm{S}_{\mathrm{L}, i}$ on a dimer as
\begin{multline}
  \mathcal{H}
  \coloneqq\frac{J_0}{2}\sum_{i=1}^L\bm{T}_{i}^2+J_1\sum_{i=1}^L\bm{T}_i\cdot\bm{T}_{i+1}\\
  +J_2\sum_{i=1}^L\bm{T}_i\cdot\bm{T}_{i+2}-h\sum_{i=1}^LT^z_i,
  \label{eq:H}
\end{multline}
where we use $\bm{T}^2_i=4+2\bm{S}_{\mathrm{U},i}\cdot\bm{S}_{\mathrm{L}, i}$ and drop the constant term.
Hereafter we focus on $\mathcal{H}$ instead of $\mathcal{H}_{\mathrm{full}}= \mathcal{H} + 2 J_0 L$.
Since each $\bm{T}_i^2$ commutes with $\mathcal{H}$, its eigenvalue $j_i(j_i+1)$ $(j_i\in\{0, 1, 2\})$ is a good quantum number.
Thus this model is equivalent to spin chains with spin quantum numbers that may vary from site to site.

\begin{figure}
  \includegraphics[width=\columnwidth]{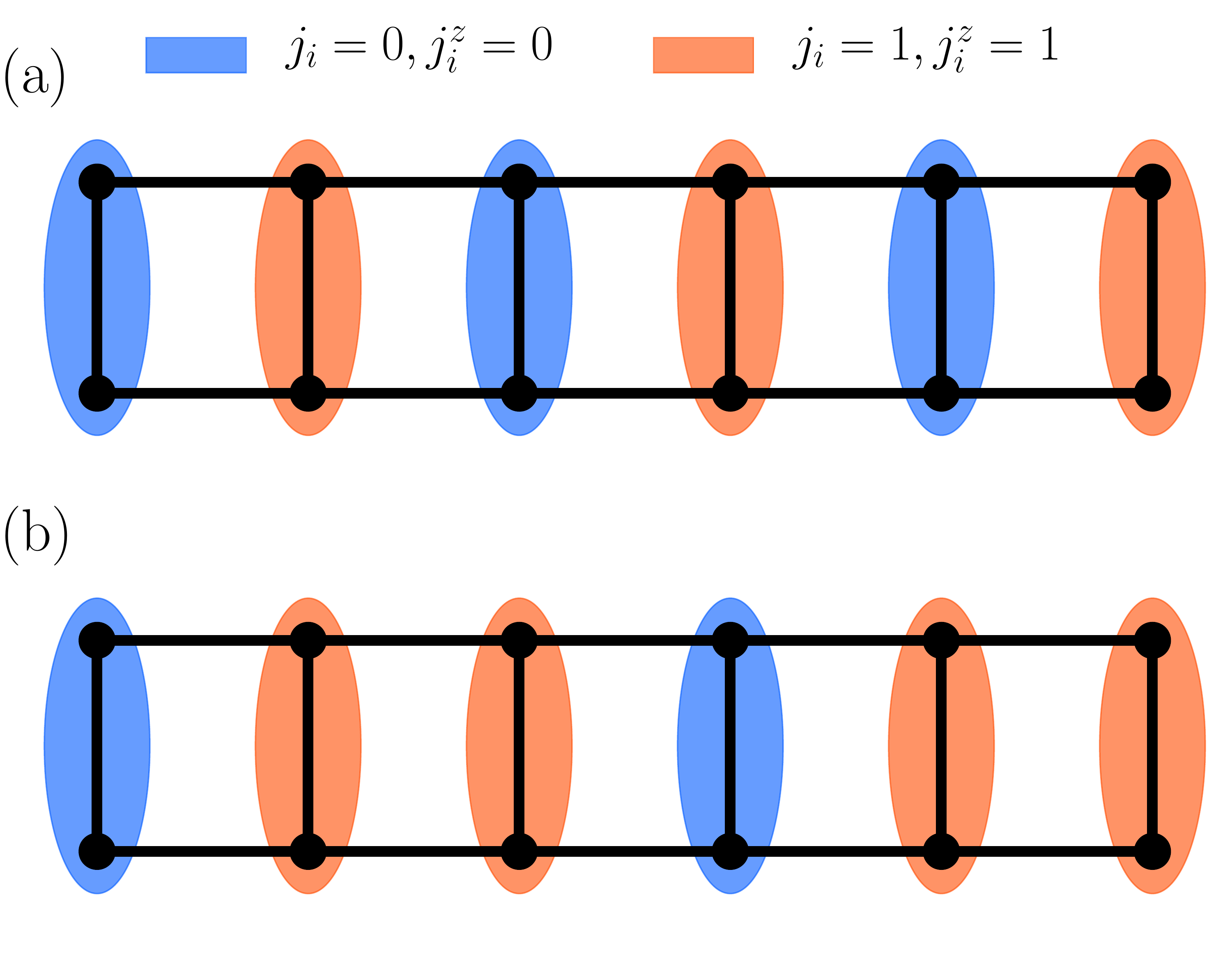}
  \caption{Schematic picture of (a) $1/4$- and (b) $1/3$-plateau states.}
  \label{fig:states}
\end{figure}

If the $J_2$ term is absent, this model is called the $S=1$ Gelfand ladder~\cite{Gelfand, HMT, ChandraSurendran}.
The $S=1/2$ Gelfand ladder has been studied by Honecker \textit{et al.}~\cite{HMT} and later, Chandra and Surendran studied the general spin case~\cite{ChandraSurendran}.
To fix our notations, we briefly review their results on the $S=1$ Gelfand ladder.
Completing the square yields the simple form of the Hamiltonian,
\begin{equation}
  \mathcal{H}=\frac{J_0-2J_1}{2}\sum_{i=1}^L\bm{T}_{i}^2+\frac{J_1}{2}\sum_{i=1}^L\left(\bm{T}_i+\bm{T}_{i+1}\right)^2-h\sum_{i=1}^LT^z_i.
\end{equation}
If $J_0\geq 2J_1 \geq 0$, which corresponds to the line $OP$ in Fig.~\ref{fig:region}, we can prove that the exact ground state is a product state,
\begin{equation}
  \Ket{\{j_i\}}\coloneqq\bigotimes_{i=1}^{L}\Ket{j_i}_i.
  \label{eq:prod_state}
\end{equation}
Here $\Ket{j_i}_i$ denotes the highest-weight state of the spin-$j_i$ representation at the $j$th rung, which is the eigenstate of $\bm{T}_{i}^2$ and $T_i^z$ with eigenvalues $j_i(j_i+1)$ and $j_i$
\footnote{Note that each $\Ket{j_i}_i$, characterized by $\bm{T}_i^2$ and $T_i^z$,
would be degenerate in general; however, we follow the notation in
Ref.~[\onlinecite{ChandraSurendran}] and do not distinguish such
states labeled by any extra quantum numbers.
}.
The ground-state energy is obtained by minimizing the energy
\begin{equation}
  \frac{J_0}{2}\sum_{i=1}^Lj_i(j_i+1)+J_1\sum_{i=1}^Lj_ij_{i+1}-h\sum_{i=1}^Lj_i\label{eq:Egel}
\end{equation}
with respect to $\{j_i\}$.
The transition fields are given as
\begin{equation}
  \begin{split}
    h_{0,1/4}&=J_0,\\
    h_{1/4,1/2}&=J_0+2J_1,\\
    h_{1/2,3/4}&=2J_0+2J_1,\\
    h_{3/4,1}&=2J_0+4J_1,
  \end{split}
  \label{eq:hc4}
\end{equation}
where $h_{m,\tilde{m}}$ is the transition field from $m$ to $\tilde{m}$ plateaus.
Starting from the zero-magnetization state at small magnetic field, the magnetization jumps to $1/4$ at $h = h_{0,1/4}$, where rung singlets ($j_{2k-1}=0$) and rung triplets ($j_{2k}=1$) are alternately aligned as depicted in Fig.~\ref{fig:states}(a).
This state has the lowest energy when $h_{0,1/4} \leq h \leq h_{1/4,1/2}$.
At $h=h_{1/4,1/2}$, the magnetization macroscopically jumps and then the $1/2$-plateau state with $j_i=1$ for all rung $i$ appears.
The $3/4$-plateau state with $(j_{2k-1}, j_{2k})=(1,2)$ becomes the ground state when $h_{1/2,3/4} \leq h \leq h_{3/4,1}$, which is sandwiched by the $1/2$-plateau state and the fully saturated state ($j_i=2$).
The $1/4$- and $3/4$-plateau states break translational symmetry and are doubly degenerate, while the $1/2$-plateau state does not.
The resulting magnetization curve at $J_1=J_0/2$ is shown as the $J_2=0$ line in Fig.~\ref{fig:mh}.

We note that the extended Gelfand ladder with $J_1=0$ is decoupled into two independent Gelfand ladders aligned alternately.
Its ground state is also a plateau state whose magnetization is a multiple of $1/4$.
The period of the ground state to be realized is different from that of the original Gelfand ladder, and a state with period four appears instead of period two.
If we consider $J_2$ as the magnitude of the nearest-neighbor interaction between rungs, the corresponding condition is $J_0\geq 2J_2 \geq 0$, which corresponds to the line $OR$ in Fig.~\ref{fig:region}.

\section{Exact solutions for plateaus at multiples of $1/6$}\label{sec:onesixth}
If $J_2=J_1/2$ and $0 \leq J_2 \leq J_0/3$, which corresponds to the line $OQ$ in Fig.~\ref{fig:region}, we can discuss the ground state of the extended Gelfand ladder by completing the square.
In this case, the ground states obtained are also product states with magnetization plateaus of multiples of $1/6$.

Let us consider a decomposition of a Hamiltonian into a sum of local Hamiltonians; $\mathcal{H}=\sum_{i=1}^{L}\mathcal{H}_i$.
In general, the sum of the lowest eigenvalues of $\mathcal{H}_i$ gives a lower bound of the total ground-state energy.
Our Hamiltonian can be decomposed as a sum of local three-rung Hamiltonians $\mathcal{H}_i$.
We will show that such a decomposition gives a product state that is the local three-rung ground state of $\mathcal{H}_i$.
Using the local ground state, we can construct an eigenstate of the total system, which is the ground state of $\mathcal{H}_i$ for all $i$.
Since such a state achieves the lower bound of the ground-state energy, it is an exact ground state of the full Hamiltonian $\mathcal{H}$.

Let us assume $J_1=2J_2$ and $0 \leq J_2 \leq J_0/3$.
Completing the square yields the simple form of a local Hamiltonian,
\begin{multline}
  \mathcal{H}_i
  = \frac{J_0-3J_2}{6}(\bm{T}_i^2+\bm{T}_{i+1}^2+\bm{T}_{i+2}^2) \\
  +\frac{J_2}{2}(\bm{T}_i+\bm{T}_{i+1}+\bm{T}_{i+2})^2 \\
  -\frac{h}{3}(T^z_i+T^z_{i+1}+T^z_{i+2}).
\end{multline}
Since all the terms in $\mathcal{H}_i$ commute with each other, the eigenvalue of $\mathcal{H}_i$ is written as
\begin{multline}
  E_i(j_i, j_{i+1}, j_{i+2},j_i^\mathrm{tot},j_i^{z, \mathrm{tot}})
  =\frac{J_0-3J_2}{6}\sum_{k=i}^{i+2} j_{k}(j_{k}+1) \\
  +\frac{J_2}{2}j_i^\mathrm{tot}(j_i^\mathrm{tot}+1)-\frac{h}{3} j_i^{z, \mathrm{tot}},
  \label{eq:E_i}
\end{multline}
where $j_i^\mathrm{tot}(j_i^\mathrm{tot}+1)$ and $j_i^{z, \mathrm{tot}}$ is the eigenvalue of $(\sum_{k=i}^{i+2} \bm{T}_k)^2$ and $\sum_{k=i}^{i+2} T_k^z$, respectively.

We can solve the problem of finding the lowest eigenvalue of $\mathcal{H}_i$ in the magnetic field by finding a minimum of Eq.~\eqref{eq:E_i} for each $j_i^\mathrm{tot}$.
The non-negativity of the field-independent terms in $E_i$ is guaranteed by $0\leq J_2 \leq J_0/3$.
Since $j_k$ is a non-negative integer and satisfies $\sum_{k=i}^{i+2} j_k \geq j_i^\mathrm{tot}$, we have
\begin{align}
  \min_{\{j_k\}} \sum_{k=i}^{i+2} 3 j_{k}(j_{k}+1) =
  \begin{cases}
    j_i^\mathrm{tot}(j_i^\mathrm{tot}+3)     & (j_i^\mathrm{tot}=3n) \\
    (j_i^\mathrm{tot}+1)(j_i^\mathrm{tot}+2) & (\text{otherwise}),   \\
  \end{cases}
\end{align}
where $n \in {0,1,2}$.
Its proof is straightforward via $3(j_i^2 + j_{i+1}^2 + j_{i+2}^2) = (j_i-j_{i+1})^2 + (j_{i+1}-j_{i+2})^2 + (j_{i+2}-j_{i})^2 + (j_i + j_{i+1} + j_{i+2})^2$.
In the case of $j_i^\mathrm{tot}=3n$, $E_i$ takes the minimum at $j_k=n$ for all $k$,
and if $j_i^\mathrm{tot}=3n\pm 1$, $(j_i, j_{i+1}, j_{i+2})=(n\pm 1, n, n)$ and its permutations achieve the minimum of $E_i$.
For the last term in Eq.~\eqref{eq:E_i}, we can set $j_i^{z, \mathrm{tot}}=j_i^\mathrm{tot}$ if the external magnetic field $h$ is positive.
As $h$ increases, $j_i^\mathrm{tot}$ rises from $0$ to $6$ in steps of one and its change occurs at
\begin{equation}
  \begin{split}
    h_{0,1/6}&=J_0,\\
    h_{1/6,1/3}&=J_0+3J_2,\\
    h_{1/3,1/2}&=J_0+6J_2,\\
    h_{1/2,2/3}&=2J_0+6J_2,\\
    h_{2/3,5/6}&=2J_0+9J_2,\\
    h_{5/6,1}&=2J_0+12J_2.
  \end{split}
  \label{eq:hc6}
\end{equation}

In all the cases discussed above, the ground state of $\mathcal{H}_i$ is
a product state because it satisfies $j_i^\mathrm{tot}=j_i^{z, \mathrm{tot}}=\sum_{k=i}^{i+2}j_k$.
In the case of $j_i^\mathrm{tot}=3n$, all dimers have the total spin $n$ and are maximally polarized.
Such a state can be extended to the whole system, which is written as
\begin{equation}
  \Ket{\psi_{3n}}= \bigotimes_{i=1}^{L} \Ket{n}_i,\label{eq:phi_3n}
\end{equation}
where $\Ket{n}_i$ denotes the highest-weight state on the rung $i$.
Obviously, it is the ground state of $\mathcal{H}_i$ for every $i$ in certain ranges of the magnetic field and achieves the lower bound of the ground-state energy of $\mathcal{H}$.
Therefore, this state is the ground state of the full Hamiltonian $\mathcal{H}$.
In the same way, we can obtain the ground state for $j_i^\mathrm{tot}=3n \pm 1$ as
\begin{equation}
  \Ket{\psi_{3n\pm 1}} = \bigotimes_{i=1}^{L/3}\Ket{n\pm 1}_{3i-2}\otimes \Ket{n}_{3i-1} \otimes \Ket{n}_{3i}.
  \label{eq:phi_3n1}
\end{equation}
This state breaks the translational symmetry and is threefold degenerate, while $\Ket{\psi_{3n}}$ is the unique ground state.

The ground state switches at the transition field Eq.~\eqref{eq:hc6}, where the magnetization macroscopically jumps.
The optimal $j_i^\mathrm{tot}$, which minimizes the energy, does not change between them, and thus the $j_i^\mathrm{tot}/6$ plateaus appear.

\section{Phase diagram of the extended Gelfand ladder in magnetic field}
\label{sec:pd}
\begin{figure}
  \includegraphics[width=\columnwidth]{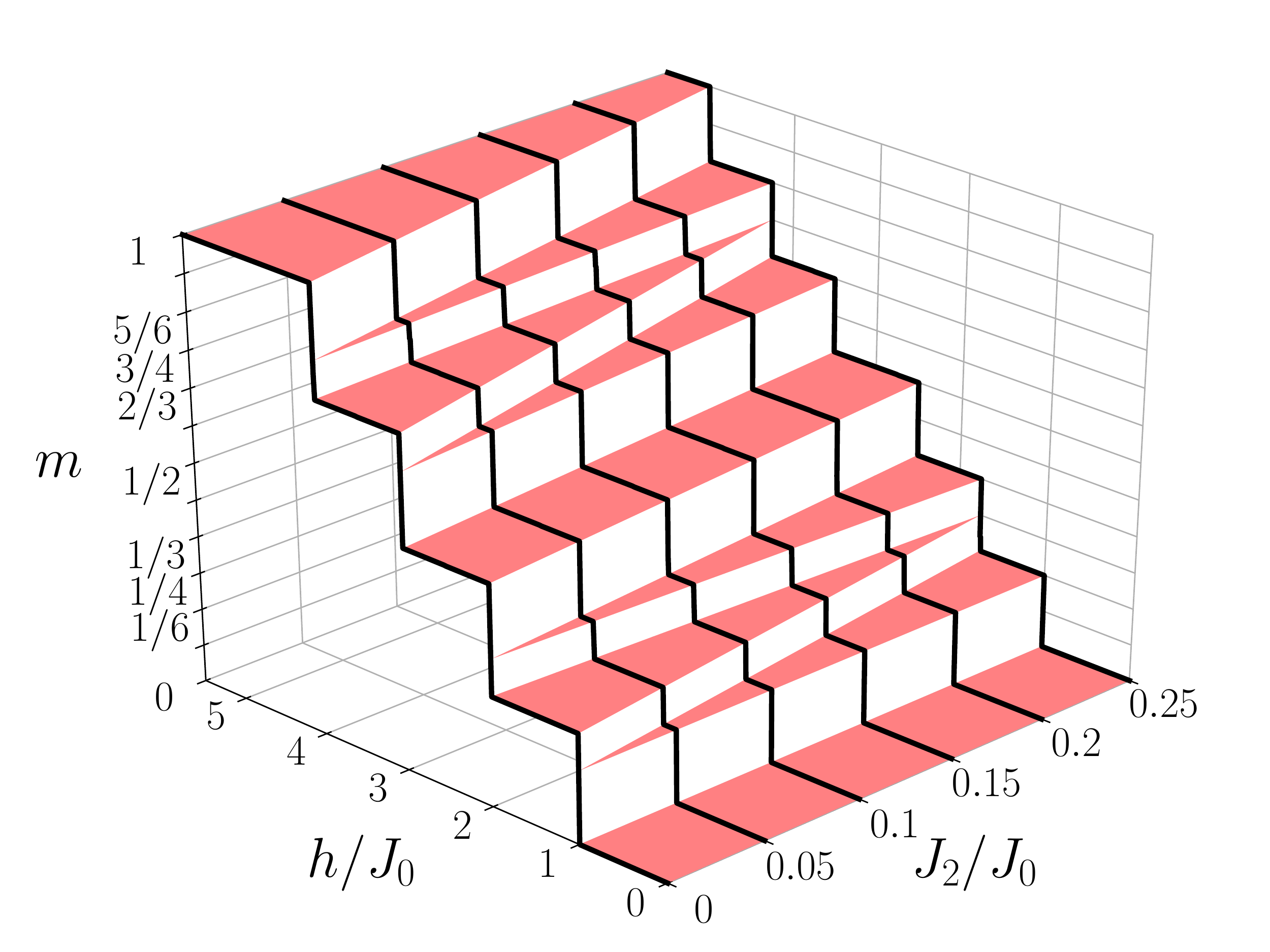}
  \caption{
    Magnetic curve with various values of $J_2$ with $J_1=J_0/2$.
    Magnetic curves suddenly jump between neighboring plateaus, and there is no nonplateau region.
    $J_2=0$ corresponds to the model considered by Chandra and Surendran~\cite{ChandraSurendran}.
    The $1/4$ and $3/4$ plateaus vanish at $J_2/J_0=1/4$.
  }
  \label{fig:mh}
\end{figure}

\begin{figure}
  \includegraphics[width=\columnwidth]{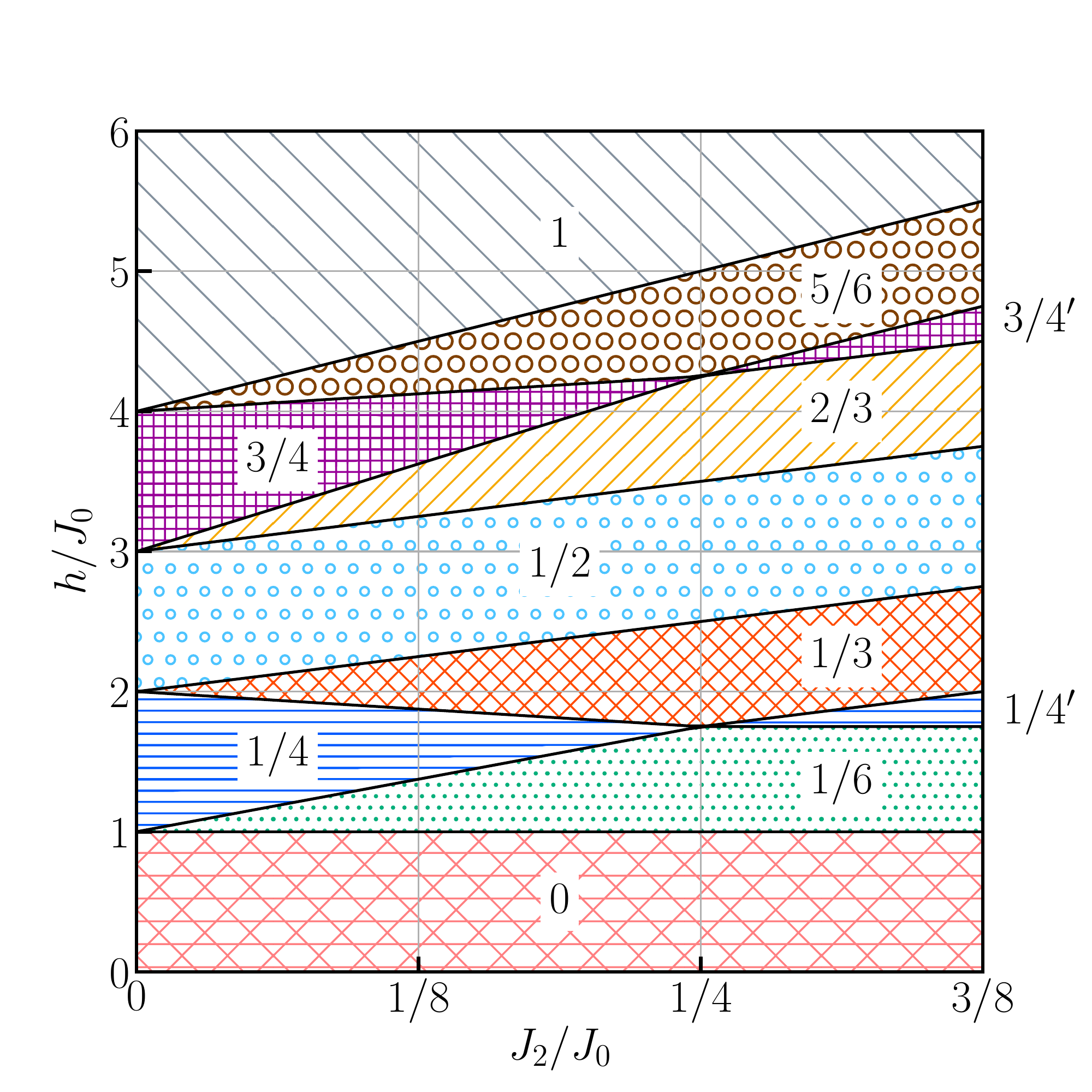}
  \caption{
    $J_2$-$h$ phase diagram at $J_1=J_0/2$.
    Numbers in the figure are values of the magnetization per site.
  }
  \label{fig:J2h}
\end{figure}

\begin{table}[b]
  \caption{Energy per rung of each plateau.
    When magnetization per site $m$ takes a value of $1/4$ or $3/4$, there are two types of ground states, period two and period four.
    The latter is denoted by $1/4'$ or $3/4'$.
  }
  \label{table:1}
  \begin{ruledtabular}
    \begin{tabular}{ccc}
      $m$    & Sequence of $\{j_i\}$ & Energy per rung         \\\hline
      $0$    & $(0)$                 & $0$                     \\
      $1/6$  & $(0,0,1)$             & $(J_0-h)/3$             \\
      $1/4$  & $(0,1)$               & $(J_0+J_2-h)/2$         \\
      $1/4'$ & $(0,0,1,1)$           & $J_0/2+J_1/4-h/2$       \\
      $1/3$  & $(0,1,1)$             & $(2J_0+J_1+J_2-2h)/3$   \\
      $1/2$  & $(1)$                 & $J_0+J_1+J_2-h$         \\
      $2/3$  & $(1,1,2)$             & $(5J_0+5J_1+5J_2-4h)/3$ \\
      $3/4$  & $(1,2)$               & $2J_0+2J_1+(5J_2-3h)/2$ \\
      $3/4'$ & $(1,1,2,2)$           & $2J_0+9J_1/4+2J_2-3h/2$ \\
      $5/6$  & $(1,2,2)$             & $(7J_0+8J_1+8J_2-5h)/3$ \\
      $1$    & $(2)$                 & $3J_0+4J_1+4J_2-2h$
    \end{tabular}
  \end{ruledtabular}
\end{table}

A straightforward calculation shows that the product state Eq.~\eqref{eq:prod_state} is an exact eigenstate of $\mathcal{H}$ with energy
\begin{multline}
  E(\{j_i\})=\frac{J_0}{2}\sum_{i=1}^L j_i\left(j_i+1\right)+J_1\sum_{i=1}^L j_i j_{i+1}\\+J_2\sum_{i=1}^L j_i j_{i+2}-h\sum_{i=1}^L j_i.\label{eq:E_tot}
\end{multline}
We will show in Appendix~\ref{sec:app_CQ} that this result can be easily obtained by organizing the Hamiltonian into terms that disappear and terms that remain when acting on the highest-weight state.

Assuming that the ground state is the product state of the form Eq.~\eqref{eq:prod_state}, the remaining problem is to find the minimum of $E(\{j_i\})$ with respect to $\{j_i\}$.
Although it is not guaranteed that the ground state is $\Ket{\{j_i\}}$, we will discuss the parameter region where the ground state is $\Ket{\{j_i\}}$ in the next section.

We calculate $E(\{j_i\})$ and summarize the energy per rung for each plateau state in Table~\ref{table:1}.
These states have trivial degeneracies that reflect translational symmetry.
We characterize these states by a sequence of $\{j_i\}$.
For example, the $1/3$-plateau state depicted in Fig.~\ref{fig:states}(b) has a period three and $(j_{3k-2}, j_{3k-1}, j_{3k})=(0, 1, 1)$.

We can determine each transition field between plateaus by comparing $E(\{j_i\})$.
The transition fields are given as
\begin{equation}
  \begin{split}
    h_{0,1/6}   &= J_0,            \\
    h_{1/6,1/4} &= J_0+3J_2,       \\
    h_{1/4,1/3} &= J_0+2J_1-J_2,   \\
    h_{1/3,1/2} &= J_0+2J_1+2J_2,  \\
    h_{1/2,2/3} &= 2J_0+2J_1+2J_2, \\
    h_{2/3,3/4} &= 2J_0+2J_1+5J_2, \\
    h_{3/4,5/6} &= 2J_0+4J_1+J_2,  \\
    h_{5/6,1}   &= 2J_0+4J_1+4J_2,
  \end{split}\label{eq:hc12}
\end{equation}
for $J_2 \leq J_1/2$.
For $0 < J_2 < J_1/2$, both series of plateaus coexist in a single magnetization curve.
When $J_2 = 0$, some transition fields take the same value (e.g., $h_{0,1/6}=h_{1/6,1/4}=J_0$ at $J_2=0$), and give results consistent with the previous work~\cite{ChandraSurendran} while the $1/4$ and $3/4$ plateaus disappear at $J_2 = J_1/2$.
Substituting $J_2 = J_1/2$ into Eq.~\eqref{eq:hc6}, we get the transition fields at $J_2 = J_1/2$ [Eq.~\eqref{eq:hc6}] obtained in Sec.~\ref{sec:onesixth}.

For $J_2 \geq J_1/2$, in contrast to the case of $J_2 \leq J_1/2$, $1/4'$ and $3/4'$ states with period four appear as $1/4$- and $3/4$-plateau states.
At $J_1=0$, the model is decoupled into two independent Gelfand ladders.
For $J_1\ll J_0$, we can understand the $1/4'$ and $3/4'$ states as the alternating arrangement of the ground states for the decoupled ladders.
The transition field slightly changes from Eq.~\eqref{eq:hc12} as
\begin{equation}
  \begin{split}
    h_{1/6,1/4'} &= J_0+\frac{3}{2}J_1,      \\
    h_{1/4',1/3} &= J_0+\frac{1}{2}J_1+2J_2, \\
    h_{2/3,3/4'} &= 2J_0+\frac{7}{2}J_1+2J_2,\\
    h_{3/4',5/6} &= 2J_0+\frac{5}{2}J_1+4J_2.
  \end{split}
\end{equation}
For $J_1/2 < J_2 \leq 3J_1/4$, both series of plateaus also coexist in a single magnetization curve.

We show obtained magnetization curves with various $J_2/J_0$ at $J_1=J_0/2$ in Fig.~\ref{fig:mh}.
At $J_2=0$, the plateaus are at $m=1/4$, $1/2$, $3/4$, and $1$, which reproduces the exact solution of the Gelfand ladder~\cite{ChandraSurendran}.
At $J_2/J_0 = 1/4$, the plateaus are at $m=1/6$, $1/3$, $1/2$, $2/3$, $5/6$, and $1$.
In the intermediate region ($0<J_2/J_0<1/4$), both $1/4$ and $1/6$ series of plateaus coexist in the magnetization curves.
Every magnetic curve suddenly jumps between neighboring plateaus, and nonplateau regions are absent.
The $J_2$-$h$ phase diagram is shown in Fig.~\ref{fig:J2h}.
For $1/4 < J_2/J_0 \leq 3/8$, the $1/4'$ and $3/4'$ states take the place of the $1/4$ and $3/4$ states.

\section{Proof that the product states are the ground states}\label{sec:ff}
We prove that the ground state of the extended Gelfand ladder Eq.~\eqref{eq:H} is a product state listed in Table~\ref{table:1} in a certain parameter region, except for degeneracy at transition fields.
To do so, we use a general argument described below~\cite{Tasaki}.

Let $\mathcal{H}^\mathrm{sum}$ be a general Hamiltonian written as a sum of subterms $\mathcal{H}^\mathrm{sub}_i$; $\mathcal{H}^\mathrm{sum}=\sum_i\mathcal{H}^\mathrm{sub}_i$.
Suppose $\Ket{\gamma}$ is the ground state of $\mathcal{H}^\mathrm{sub}_i$ for all $i$, i.e.,
$\mathcal{H}^\mathrm{sub}_i\Ket{\gamma}=E^\mathrm{GS}_i\Ket{\gamma}$ where $E^\mathrm{GS}_i$ is the ground-state energy of $\mathcal{H}^\mathrm{sub}_i$.
Then $\Ket{\gamma}$ is also the ground state of $\mathcal{H}^\mathrm{sum}$, and its energy is $\sum_iE^\mathrm{GS}_i$.
If a Hamiltonian is written as $\mathcal{H}^\mathrm{sum}$, and a state $\Ket{\gamma}$ which satisfies the above property exists, the Hamiltonian is called frustration-free.

Hereafter, we consider a Hamiltonian consisting of two terms parametrized by $0\leq t \leq 1$ as
\begin{align}
  \mathcal{H}^\mathrm{FF}(t) & \coloneqq(1-t)\mathcal{H}(J_1,J_2,h)+t\mathcal{H}(J_1',J_2',h') \\
                             &
  \begin{multlined}
    =\frac{J_0}{2}\sum_{i=1}^L\bm{T}_{i}^2+J_1(t)\sum_{i=1}^L\bm{T}_i\cdot\bm{T}_{i+1}\\
    +J_2(t)\sum_{i=1}^L\bm{T}_i\cdot\bm{T}_{i+2}-h(t)\sum_{i=1}^LT^z_i,
  \end{multlined}
\end{align}
where $\mathcal{H}(J_1,J_2,h)$ is a Hamiltonian of the extended Gelfand ladder Eq.~\eqref{eq:H} with certain parameters $J_1,J_2$, and $h$.
$J_0$ is the same for both $\mathcal{H}(J_1,J_2,h)$ and $\mathcal{H}(J_1',J_2',h')$.
$J_i(t)\ (i=1, 2)$ and $h(t)$ are defined as $J_i(t)\coloneqq(1-t)J_i+tJ_i'$ and $h(t)\coloneqq(1-t)h+th'$.
Such a parametrized Hamiltonian $\mathcal{H}^\mathrm{FF}(t)$ is a realization of the extended Gelfand ladder Eq.~\eqref{eq:H}.
If we rewrite $\mathcal{H}^\mathrm{sub}_1=(1-t)\mathcal{H}(J_1,J_2,h)$ and $\mathcal{H}^\mathrm{sub}_2=t\mathcal{H}(J_1',J_2',h')$, $\mathcal{H}^\mathrm{FF}(t)$ is written as a sum of subterms.

Take $J_1$ and $J_2$ along with $J_1'$ and $J_2'$ on two different lines chosen from among $OP$, $OR$, and $OQ$ in Fig.~\ref{fig:region}, where the exact ground state has been already known.
Now, we choose $h$ and $h'$ such that $\mathcal{H}(J_1, J_2, h)$ and $\mathcal{H}(J_1', J_2', h')$ share the ground state.
The interpolating Hamiltonian $\mathcal{H}^\mathrm{FF}(t)$ thereby becomes frustration-free, and the shared ground state is also the ground state of $\mathcal{H}^\mathrm{FF}(t)$.

For example, take $J_1$ and $J_2$ on $OP$, and $J_1'$ and $J_2'$ on $OQ$, respectively, and consider interpolating the straight line with slope two, then $J_1(t)=(1+t/3)J_1$ and $J_2(t)=2tJ_1/3\ (0\leq J_1\leq J_0/2)$.
We consider the $1/4$-plateau state $\Ket{\gamma}=\Ket{(0,1)}$ as an example.
Take $h$ that satisfies $h_{0,1/4}\leq h \leq h_{1/4,1/2}$, and $h'=h_{1/6,1/3}$, then $\Ket{(0,1)}$ is the ground state of $\mathcal{H}^\mathrm{FF}(t)$ from the above argument.
In this way, the product states with the magnetization of multiples of $1/4$ degenerate with those of $1/6$ at the transition fields on $OQ$.
Similarly, the product states with the magnetization of multiples of $1/6$ degenerate with those of $1/4$ at the transition fields on $OP$ and $OR$.
Therefore, during the sweep of $h$, we can always take $h'$ such that the same state as the ground state of $\mathcal{H}(J_1,J_2,h)$ realizes the ground state of $\mathcal{H}(J_1',J_2',h')$.

Another example is an interpolation between $OQ$ and $OR$.
Take $J_1$ and $J_2$ on $OQ$, and $J_1'$ and $J_2'$ on $OR$, respectively,
Suppose $J_1$ and $J_2$ are on $OQ$ along with $J_1'$ and $J_2'$ are on $OR$, and consider interpolating the straight line with slope $-1/4$.
In this case, we have $J_1(t)=(1-t)J_1$ and $J_2(t)=(1/2+t/4)J_1\ (0\leq J_1\leq 2J_0/3)$, then the $i/4'$ and $j/6$ states appear as the ground states.

The parameter region where its ground state can be determined by drawing a line is made by sweeping a line across two lines among $OP$, $OQ$, and $OR$ and hence corresponds to the colored region in Fig.~\ref{fig:region}.
The former example corresponds to the lower red region, and the latter corresponds to the upper blue region.

By using this construction, we can draw the phase diagram without the energy comparison as we performed in Sec.~\ref{sec:pd}.
In Appendix~\ref{sec:app_CQ}, we will show that the colored region is equivalent to a positive semidefinite condition for coefficients of a decomposed Hamiltonian.

\section{Conclusion}
\label{sec:sum}
We examined the possibility of realizing several series of magnetization plateaus in a simple Heisenberg model by taking a ladder system for $S=1$ as an example.
Motivated by the exactly solvable $S=1$ Heisenberg ladder that exhibits plateaus at multiples of $1/4$, we constructed a new solvable model, the extended Gelfand ladder, that holds plateaus at multiples of $1/6$.
We have also shown that the product states are the ground states of the extended Gelfand ladder in the colored region shown in Fig.~\ref{fig:region}.
As in the previous work \cite{ChandraSurendran}, the result can be generalized for general spin $S$.
To search for coexisting magnetic plateaus, we investigated the phase diagram by calculating the energies of the product states.
We confirmed that the magnetization plateaus of both series appear with a small next-nearest neighbor interaction.

The magnetization curve always exhibits sudden jumps between neighboring plateaus as shown in Figs.~\ref{fig:mh} and \ref{fig:J2h}.
A similar jump has been found in the previous studies of the Gelfand ladder~\cite{HMT, ChandraSurendran} and in the Heisenberg antiferromagnet on the kagome lattice~\cite{Schulenburg, Nishimoto, Okuma}.
These jumps have the same origin; the excited states are also product states, and quasiparticles are localized with a flat band formed, although the types of the excited product states and therefore those of the quasiparticles depend on the phase.
In the Heisenberg antiferromagnet on the kagome lattice, some magnons are localized, and such a jump is just below the saturated magnetization.
On the other hand, in the Gelfand ladder and our extended Gelfand ladder, excitations that change $\bm{T}^2_i$ are localized, and the jumps always occur between the plateaus.

The absence of nonplateau regions in the present model results from the special condition that the magnitude of the interaction in the leg direction is equal to that of the diagonal direction over different rungs and legs.
With this condition, the Hamiltonian can be written in a spin $\bm{T}_i=\bm{S}_{\mathrm{U},i} + \bm{S}_{\mathrm{L},i}$ rather than in $\bm{S}_{\mathrm{U(L)},i}$, and $\bm{T}_i^2$ becomes a good quantum number for each $i$.
As a result, many low-energy eigenstates would be product states.
The energy gap of the excitation from the ground state to the lower excited product state would be smaller than that of the magnon excitation.

In real ladder materials, the interaction strength would likely be different between the leg and diagonal directions.
Hence, $\bm{T}_i^2$ would not be a good quantum number, and sudden jumps would turn into gradually-increasing regions.
Indeed, a previous study has shown that changing the leg and diagonal interactions allows magnon excitations and results in a gradual increase of the magnetization~\cite{Michaud}.
We note, however, that the plateaus do not immediately disappear with an infinitesimal perturbation.

A key feature of the mechanism to form magnetization plateaus is rung dimerization.
The ground states should be the product states of rung dimers in our extended Gelfand ladder, which is an ideal situation.
We have determined the dimer structure of the plateau states, e.g., the obtained $1/3$-plateau state is the singlet-triplet-triplet state.
Experiments that provide information on local spin states, such as magnetic resonance experiments, would verify our prediction.

The extended Gelfand ladder shows the plateaus at multiples of $1/6$, whereas the $1/6$ plateau is not observed in BIP-TENO experiment~\cite{Nomura}.
For the case of plateaus at multiples of $1/4$, the $1/4$ plateau on the low-field side is likely to be less stable than the $3/4$ plateau~\cite{Michaud}.
Violation of the solvability condition perhaps suppresses the 1/6 plateau more than the others.
Numerical investigation of the system perturbed away from the solvable limit remains for future work.

For the magnetization plateau to be realized in the BIP-TENO by the mechanism we have shown, the effective model of the BIP-TENO must be close to the extended Gelfand ladder.
While the latter includes the farther neighbor interactions $J_2$ and $J_3$, they are usually much smaller than the nearest-neighbor interaction in inorganic solids.
However, since the size of the BIP-TENO molecule is comparable to the distance between two sites connected by $J_3$, it may not be too outrageous to assume a non-negligible amplitude of $J_3$.
The previous \textit{ab initio} and experimental studies did not take into account farther neighbor interactions~\cite{Katoh1,Taniguchi}.
It would be helpful to estimate the amplitude of these farther neighbor interactions by an \textit{ab initio} calculation. \acknowledgments
The authors wish to thank S. C. Furuya, Y. H. Matsuda, A. Ikeda, and M. Ohno for helpful discussions and comments.
This research was supported by JSPS KAKENHI Grant No. JP19H01809 and No. JP20K03780.
H. Katsura was supported in part by JSPS Grant-in-Aid for Scientific Research on Innovative Areas, Grant No. JP20H04630, JSPS KAKENHI Grant No. JP18K03445, and the Inamori Foundation.

\appendix*
\section{Region of positive semidefinite condition}\label{sec:app_CQ}
In this Appendix, we will see that a certain positive semidefinite condition corresponds to the colored region of Fig.~\ref{fig:region}, where the product state Eq.~\eqref{eq:prod_state} is the ground state of the extended Gelfand ladder.
We will also give another derivation of Eq.~\eqref{eq:E_tot}, the energy of the product state Eq.~\eqref{eq:prod_state}.

We consider general decomposition of the Gelfand ladder Eq.~\eqref{eq:H} as
\begin{multline}
  \mathcal{H}_\mathrm{A}\coloneqq
  A\sum_{i=1}^L\bm{T}_i^2
  +B\sum_{i=1}^L(\bm{T}_i+\bm{T}_{i+1})^2
  +C\sum_{i=1}^L(\bm{T}_i+\bm{T}_{i+2})^2\\
  +D\sum_{i=1}^L(\bm{T}_i+\bm{T}_{i+1}+\bm{T}_{i+2})^2
  -h\sum_{i=1}^LT_i^z,\label{eq:H_A}
\end{multline}
where the four parameters satisfy
\begin{align}
  J_0 & = 2A + 4B + 4C + 6D, \\
  J_1 & = 2B + 4D,           \\
  J_2 & = 2C + 2D.
  \label{eq:ABCD_to_J}
\end{align}
Introducing the ladder operators, $T^\pm_i\coloneqq T_i^x\pm\mathrm{i}T_i^y$ and using identities like $\bm{T}_i^2=T_i^-T_i^++T_i^z\left(T^z_i+1\right)$, we can decompose Eq.~\eqref{eq:H_A} into two parts; $\mathcal{H}_\mathrm{A}=\mathcal{H}_\mathrm{C}+\mathcal{H}_\mathrm{Q}$.
The ``classical'' part denoted by $\mathcal{H}_\mathrm{C}$ only contains $T_i^z$ operators,
\begin{align}
  \mathcal{H}_\mathrm{C} & \coloneqq A\sum_{i=1}^LT_i^z\left(T_i^z+1\right)\notag                                                    \\
                         & \quad+B\sum_{i=1}^L\left(T_i^z+T_{i+1}^z\right)\left(T_i^z+T_{i+1}^z+1\right)\notag                       \\
                         & \quad+C\sum_{i=1}^L\left(T_i^z+T_{i+2}^z\right)\left(T_i^z+T_{i+2}^z+1\right)\notag                       \\
                         & \quad+D\sum_{i=1}^L\left(T_i^z+T_{i+1}^z+T_{i+2}^z\right)\left(T_i^z+T_{i+1}^z+T_{i+2}^{z}+1\right)\notag \\
                         & \quad-h\sum_{i=1}^LT_i^z                                                                                  \\
                         & =\frac{J_0}{2}\sum_{i=1}^LT_i^z\left(T^z_i+1\right)+J_1\sum_{i=1}^LT_i^zT_{i+1}^z\notag                   \\
                         & \quad+J_2\sum_{i=1}^LT_i^zT_{i+2}^z-h\sum_{i=1}^LT^z_i,
  \label{eq:H_C}
\end{align}
whereas the ``quantum'' part denoted by $\mathcal{H}_\mathrm{Q}$ is given by
\begin{align}
  \mathcal{H}_\mathrm{Q}
   & \coloneqq A\sum_{i=1}^LT_i^-T_i^++B\sum_{i=1}^L\left(T_i^-+T_{i+1}^-\right)\left(T_i^++T_{i+1}^+\right)\notag \\
   & \quad+C\sum_{i=1}^L\left(T_i^-+T_{i+2}^-\right)\left(T_i^++T_{i+2}^+\right)\notag                             \\
   & \quad+D\sum_{i=1}^L\left(T_i^-+T_{i+1}^-+T_{i+2}^-\right)\left(T_i^++T_{i+1}^++T_{i+2}^+\right).
  \label{eq:H_Q}
\end{align}
In $\mathcal{H}_\mathrm{Q}$, the raising operators $T^+_i$ always stand to the right of the lowering operators $T^-_i$ so that they annihilate the state $\Ket{j_i}_i$.
This rewriting results in $\mathcal{H}_\mathrm{Q}\Ket{j_i}_i=0$ and hence $\Braket{\{j_i\}|\mathcal{H}_\mathrm{A}|\{j_i\}} =\Braket{\{j_i\}|\mathcal{H}_\mathrm{C}|\{j_i\}}$.
Therefore, just replacing $T^z_i$ to $j_i$ in Eq.~\eqref{eq:H_C}, we obtain Eq.~\eqref{eq:E_tot}.

The positive semidefinite condition
\begin{equation}
  A, B, C, D\geq 0\label{eq:gene_pd}
\end{equation}
restricts the range of interactions.
With some arithmetic, the conditions~\eqref{eq:ABCD_to_J} and \eqref{eq:gene_pd} are transformed into the following ones.
\begin{equation}
  \begin{gathered}
    J_0\geq 0, \quad J_1 \geq 0, \quad J_2\geq 0, \\
    J_0 \geq 2J_1 - J_2, \quad J_0 \geq \frac{J_1}{2} + 2J_2,
  \end{gathered}
  \label{eq:pos_cond}
\end{equation}
which is the same as the colored region in Fig.~\ref{fig:region}.

We note that $\mathcal{H}_\mathrm{C}$ is related to a Hamiltonian called the frustrated Ising-Heisenberg ladder~\cite{Strecka1, Strecka2}.
The slight difference is that their model includes intrarung exchange interactions.
In both cases, we can evaluate, thanks to its classical nature, finite-temperature quantities of $\mathcal{H}_\mathrm{C}$ by using the transfer matrix method~\cite{Strecka2}.

\bibliography{gelfand}
\end{document}